# Comment on 'Field ion microscopy characterized tips in noncontact atomic force microscopy: Quantification of long-range force interactions'

William Paul and Peter Grütter

*Department of Physics, Faculty of Science, McGill University, Montreal, Canada.*

A recent article by Falter *et al.* (Phys. Rev. B 87, 115412 (2013)) presents experimental results using field ion microscopy (FIM) characterized tips in noncontact atomic force microscopy in order to characterize electrostatic and van der Waals long range forces. In the article, the tip radius was substantially underestimated at ~4.7 nm rather than ~8.0 nm due to subtleties in the application of the ring counting method. We point out where common errors in ring counting arise in order to benefit future experimental work in which the determination of tip radius by FIM is important.

Recently, Falter *et al.* reported the use of a field ion microscope (FIM) to characterize tips implemented in an atomic force microscope (AFM) employing a qPlus sensor [1]. Their work represents a significant advance in the implementation of FIM tips in scanning probe microscopy, and they present a novel way to determine electrostatic and van der Waals long range forces in these measurements based on the tip shape. The radius that the authors extract from the ring counting method was underestimated at ~4.7 nm and should be ~8.0 nm due to improper ring counting. The ring counting method was discussed in the appendix of our recent publication on the implementation of FIM tips in SPM [2], where some factors leading to underestimated tip radii were presented. Since Falter *et al.* show that important applications of FIM tips in SPM are on the horizon, we feel the need to clarify the use of ring counting in order to properly determine tip radii.

We note that some of the concerns raised here have appeared previously in the work of Webber on ring counting [3] and on the detailed determination of tip shape [4]. In this comment, we present ball models of tungsten tip apices having the same crystallographic orientation as the tip presented by Falter *et al.* and demonstrate where two common errors arise in radius determination. The first common error occurs when visible rings in FIM correspond to more than one atomic plane. The second common error, which we cover here for completeness (it does not appear in the work of Falter *et al.*) is the improper choice of interlayer plane spacing.

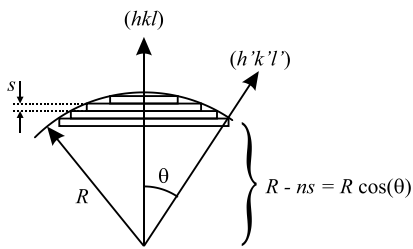

Figure 1: Side view of a tip with radius R showing the geometry of ring counting. Planes with spacing *s* are normal to the *(hkl)* direction, and *n* steps of height *s* are counted from pole *(hkl)* to *(h'k'l')*.

Ring counting is conventionally, but somewhat misleadingly, described in the following way [5]: Assuming a spherical envelope of the tip apex, the local radius of curvature between two crystallographic poles separated by angle $\theta$ is determined by the number of rings $n$ between them and the interlayer spacing $s$ corresponding to the appropriate crystal plane:

$$R = \frac{ns}{1 - \cos\theta} \qquad (1)$$

Eq. (1) describes the geometry illustrated in Figure 1.

For the ring counting estimations to be accurate, each counted ring must correspond to a *single* atomic step of type *(hkl)*. We will soon demonstrate that in FIM micrographs, a single 'ring' can correspond to multiple atomic planes of height *s* – counting rings is not equivalent to counting steps. The underestimation of FIM tip radii occurs when the rings appearing in the micrograph correspond to *more than one* atomic plane. We suggest that the standard description of ring counting could be more precisely expressed as: Assuming a spherical envelope of the tip apex, the local radius of curvature is determined by counting the number $n$ of steps of height $s$ between crystallographic poles with angle $\theta$ between them.

The ring counting method can be applied to different orientations of *(hkl)* pole, but the rings of the (110) plane are the most straightforward to identify because they have the largest step height in the bcc crystal. The (110) plane also happens to be the apex of most polycrystalline tungsten tips because of the crystallographic texture due to the wire manufacturing by cold drawing [6].

In Figure 2, we present ball models of tungsten tip apices of radii 3.0, 4.7, 6.0, and 9.0 nm. The models were created by carving a hemispherical shell from a bcc crystal with a (110) apex. The atoms in the outermost 0.05 nm shell are shaded lighter in order to highlight the atoms at terrace edges which would be imaged brightly in FIM (a common method of visualizing atomic geometry with ball models [7,8]).

As the tip radius increases, the size of crystallographic facets increases – for example, the (211) facet has just two rows of atoms in Figure 2(a), but has five rows of atoms in Figure 2(c). Correspondingly, the number of rings increases between the centers of crystallographic poles. From Eq. (1), we have calculated the expected number of rings *n* between the (110) apex and the (111) and (211) poles (black and white circles, respectively). The center of these planes is indicated by the circles on the ball models.

The expected number of rings corresponds well with the number of rings counted from the tip apex to the center of the crystallographic poles, indicated by quarter circles at the edge of the (110) steps. The number of rings can be thought of as the number of (110) steps that must be descended from the (110) apex in order to reach the center of the *(h'k'l')* pole in question. For these tips of relatively small radii and for the small angles between the apex and the (111) and (211) directions, there is a single (110) plane for each ring, therefore the estimation of ring counting is accurate.

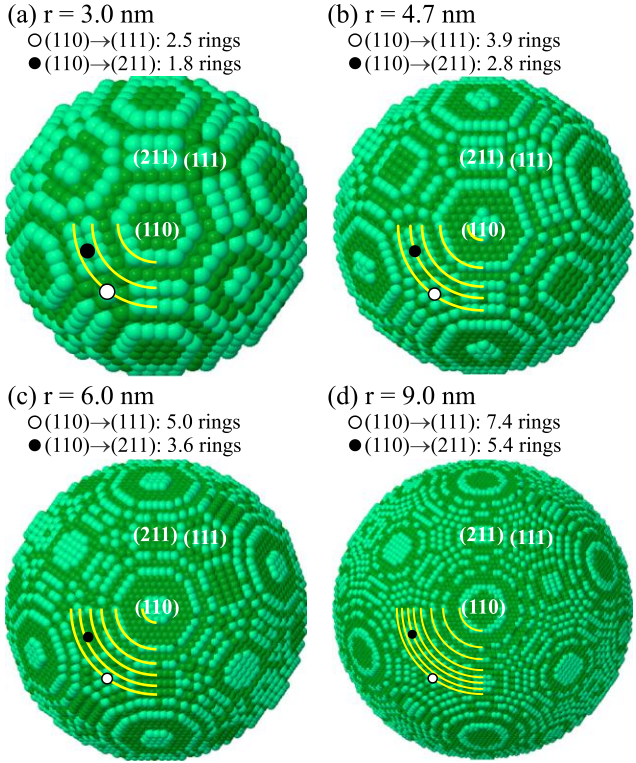

Figure 2: Top view (looking toward the apex) of ball models of W(110) tips with radii (a) 3.0 nm, (b) 4.7 nm (the radius reported by Falter *et al*.), (c) 6.0 nm, and (d) 9.0 nm.

Comparing the expected number of rings to the rings counted on the ball models, a better estimation of the tip radius is obtained using a ring count of $n - 1$ instead of $n$: this off-by-one error can be explained in Figure 1 where it is the number of steps crossed to get from *(hkl)* to *(h'k'l')* that counts (one less than the total number of rings). This is a small correction, however it should be considered in future work, especially with small radius tips. This concern is equivalently expressed by starting to count rings at 0 rather than 1, as done by Webber [3].

The first serious error that occurs in ring counting is that of counting FIM rings which correspond to more than one atomic plane of type *(hkl)*. This is problematic in the first 4 rows of Table 1 in Falter *et al*., where the authors estimate the radius between (110)-type poles which are 60° apart. Shown in Figure 3 is the side view of the 6.0 nm tip presented in Figure 2(c). Counting the rings between (110) and (01$\bar{1}$) yields 9 rings, indicated by red arrows on the right half of the image. As the angle from the apex increases, the counted rings correspond to more than one (110) plane. This is illustrated by the blue "ruler" on the left half of the image where we have counted 15 atomic planes between the (110) apex and the center of the (101) plane. Using $n = 9$, we obtain a radius of 4.0 nm, whereas a more accurate estimate is obtained using $n - 1 = 14$, yielding a radius of 6.2 nm.

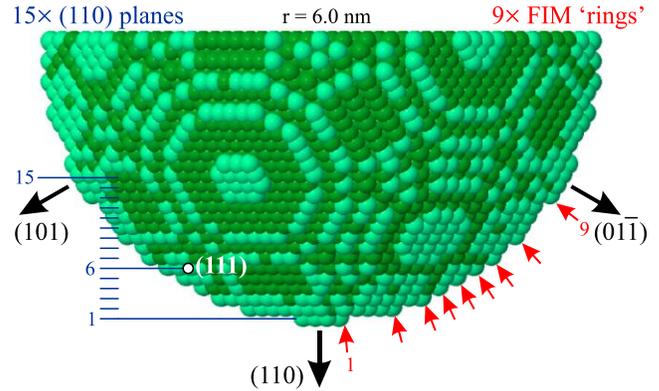

Figure 3: Side view of the 6.0 nm radius W(110) tip presented in Figure 2(c). There are 15 layers of (110) planes between the apex and the (101) facet (dark blue lines), but only 9 'rings' apparent in the micrograph (red arrows). The estimation based on 14 × 0.446 nm yields a radius of 6.2 nm, whereas 9 × 0.446 nm yields an underestimated radius of 4.0 nm. The (111) apex is shown at layer #6 from the (110) apex.

Figure 3 demonstrates that in the vicinity of the (01$\bar{1}$) pole, the (110) plane edges form the smooth surface of (01$\bar{1}$) planes and are therefore not imaged in FIM. This effect becomes more problematic as tip radius increases and as the angle between *(hkl)* and *(h'k'l')* increases. Even though ring counting between (110) and (211) poles was robust in Figure 2, for large tips of radius ~30 nm, such as the one presented by Urban *et al*. [9], the (110) plane edges become difficult to count individually as they are consumed by the relatively flat (211) planes.

The second common ring counting error, which does not appear in Falter *et al*., is the fact that the interplanar spacing *s* must correspond to the *(hkl)* pole – the initial pole – not *(h'k'l')*, the final pole. Using the (111) plane spacing rather than (110), Urban *et al*. [9] deduced a tip radius of 14.4 nm (in combination with the difficulty in counting individual planes on a large tip), whereas we estimate a radius of the order ~32 nm by counting ~21 rings from (110) to (211). These problems are again seen in Pitters *et al*. [10] where the 3 and 5 nm radii tips determined by the authors are actually of the order 6 and 10 nm, and also in Rezeq *et al*. [11] where the 1.4 nm radius apex should be ~4.8 nm (although the image resolution makes identifying the (110) and (111) planes difficult).

By carrying out careful ring counting, we estimate that the tip used by Falter *et al*. has a radius of 8.0 ± 0.8 nm rather than the reported 4.7 ± 1.1 nm. Table 1 reports the local radii of curvature we obtain near the (110) apex of the tip used by Falter *et al*. A final point of discussion is what uncertainty to attribute to this value: an important consideration is that since the radius of curvature is *local* (there is no reason that the spherical envelope should be perfect everywhere), we cannot expect values to converge to some 'true' value. In this case, the standard deviation of the obtained values expresses the magnitude of deviation from a spherical envelope. The reported uncertainty could also be based on counting integer numbers of rings (in this case, the maximum possible error of one ring gives the $\Delta R$ value used by Falter *et al.*). The final choice of which type of uncertainty to report depends on the context of the experiment.

Table 1: Local radius of curvature determined by counting $n – 1$ rings from the [110] apex to various $[h'k'l']$ poles. Here we ensure that the $[h'k'l']$ poles chosen are within a small enough angle from the apex that single steps are being counted. The 8 local radii we determine have an average value of $\langle R_{apex} \rangle = 8.0 \pm 0.8$ nm, where this uncertainty corresponds to the standard deviation of the obtained values. Identification of the $[h'k'l']$ poles can be done by comparing the planes visible in the FIM image to the low-index poles identified in a stereographic projection map of a bcc (110) crystal (see , for example, in Ref [5]).

| $R_{[110]\#n-1[h'k'l']}$ | Local curvature (nm) | $\Delta R$ in $(h'k'l')$ direction (nm) |
|---|---|---|
| $R_{[110]\#4[211]}$ | 6.68 | ± 1.67 |
| $R_{[110]\#n[121]}$ | 8.35 | ± 1.67 |
| $R_{[110]\#n[12\bar{1}]}$ | 8.35 | ± 1.67 |
| $R_{[110]\#n[21\bar{1}]}$ | 8.35 | ± 1.67 |
| $R_{[110]\#n[111]}$ | 7.26 | ± 1.21 |
| $R_{[110]\#n[11\bar{1}]}$ | 7.26 | ± 1.21 |
| $R_{[110]\#n[31\bar{1}]}$ | 9.06 | ± 1.51 |
| $R_{[110]\#n[13\bar{1}]}$ | 9.06 | ± 1.51 |
| $\langle R_{apex} \rangle$ | 8.0 ± 0.8 | ± 1.52 |

To summarize, the key elements to proper radius determination by ring counting are as follows:

- $n$ must be accurately determined – its value must correspond to the number *single* steps of type $s$
- $s$ must correspond the plane spacing of *(hkl)*, not *(h'k'l')*
- $n$ should be replaced by $n – 1$ for a more accurate radius estimation for small tips (or equivalently, counting should start at zero)

We also urge the use of (110) planes due to their large step height and not (111); due to the small (111) step height, rings corresponding to single (111) steps are very difficult to discern. The routine construction of ball models of FIM tips is also helpful to familiarize the experimenter with the appearance of tips with different radii.

We look forward to future fruitful experimental results using FIM tips in SPM experiments and hope that a careful consideration of ring counting will aid in the accurate determination of tip radii.

The authors acknowledge NSERC, CIFAR and RQMP for funding